\documentclass[aps,prl,twocolumn,preprintnumbers,
superscriptaddress,floatfix]{revtex4}

\usepackage{amssymb,multirow,amsmath}
\usepackage{graphicx,color}

\begin{document}

\newcommand{\Tr}{\mbox{Tr\,}}
\newcommand{\beq}{\begin{equation}}
\newcommand{\eeq}{\end{equation}}
\newcommand{\bea}{\begin{eqnarray}}
\newcommand{\eea}{\end{eqnarray}}
\renewcommand{\Re}{\mbox{Re}\,}
\renewcommand{\Im}{\mbox{Im}\,}

\title{Thermal Transitions in Domain Wall AdS/QCD}

\author{Nick Evans}
\affiliation{ STAG Research Centre \&  Physics and Astronomy, University of
Southampton, Southampton, SO17 1BJ, UK}

\author{Jack Mitchell}
\affiliation{ STAG Research Centre \&  Physics and Astronomy, University of
Southampton, Southampton, SO17 1BJ, UK}

\begin{abstract}
We study thermal transitions in a Domain Wall AdS/QCD model. The model is based on the D5/probe D7 system with a discontinuous mass profile which restricts chiral fermions to 3+1 dimensional domain walls. Fluctuations on the domain wall are dual to the quark mass and condensate and reveal the relation between domain wall separation and the quark mass. The massive quarks exhibit a second order thermal, meson melting transition. Witten's multi-trace prescription can be used to interpret these configurations as having a dynamical mass from a Nambu-Jona-Lasinio interaction - here the transition is first order. Confinement can be introduced into the gauge sector by compactifying one direction of the D5. Compactification induces chiral symmetry breaking and there is a first order thermal restoration transition. If an NJL interaction is also introduced then the confinement and chiral symmetry breaking scales can be separated. 
\end{abstract}

\maketitle
In this paper we will continue our investigation of domain wall fermions in holography as discussed in \cite{CruzRojas:2021pql,Evans:2021zzm}. The domain wall trick \cite{Kaplan:1992bt} is now a standard piece of lattice technology for studying chiral fermions. In holography \cite{Maldacena:1997re}, realizing chiral fermions should also allow the construction of duals for a wider set of theories that includes QCD and beyond. 

In \cite{CruzRojas:2021pql} we studied a basic domain wall construction in the well known D3/probe D7 system \cite{Karch:2002sh}
that describes quarks in ${\cal N}=4$ Super Yang-Mills theory. A quark mass that is spatially dependent in one of the directions of the 3+1 dimensional theory was introduced. If the mass, $m$, sharply switches sign from a large positive value to an equal negative value then that marks the position of the domain wall.  The (one) lower dimension fermions are localized on the  $m=0$ contour, which can be found in the AdS space. We found U-shaped configurations linking two neighbouring domain walls. The D7 world-volume field that is dual to the quark mass and condensate can take non-zero values on the domain wall at next order in the computation. We solved for this field setting the IR boundary condition from the radial position of the tip of the U-shape. That  demonstrated that the mass of the quarks, in that example, is inversely proportional to the separation of the domain walls. We will see generically, that this story is more complicated. In theories with multiple scales, such as those described by a UV cut off and a temperature, the width to mass identification is more complex. The ability to explicitly describe the quark mass and condensate via this field is one of the benefits of this methodology. For the D3/probe D7 system the fermions live in 2+1 dimensions and are not chiral.

A holographic description of chiral fermions in 3+1 dimensions was constructed in \cite{Evans:2021zzm} where we used the D5/probe D7 system in the supersymmetry preserving configuration

     \begin{center}
 \begin{tabular}{c|c c c c c c c c c c}
      &0&1&2&3&4&5&6&7&8&9   \\   \hline
      D5&-&-&-&-&-&(-)&$\bullet$ &$\bullet$ &$\bullet$ &$\bullet$ \\
      D7&-&-&-&-&-&$\bullet$  &-&-&- &$\bullet$ 
 \end{tabular}
\end{center} \vspace{-1cm} 

\beq \label{one}\eeq

The $x_4$ direction is marked in parentheses because it can be compactified to introduce confinement in the gauge theory on the D5 brane. This was the main focus of \cite{Evans:2021zzm} where we showed that the confinement led to chiral symmetry breaking and we studied the QCD-like meson spectrum.

In this paper we wish to study this system further by examining its thermal transitions. The system can display meson melting, chiral restoration and deconfinement transitions and it is important to map out these behaviours. The D5/probe D7 system is known to have an ill behaved UV - the coupling of the system grows into the UV and in \cite{Myers:2006qr} it was shown that the mesons of the basic 4+1d system are not normalizable. However, the domain wall system seems to avoid these inconsistencies because it is dimensionally reduced. We will see a further sign of the odd UV behaviour in the uncompactified model - all connected domain wall pairs asymptote to the same UV separation no matter the quark mass they describe (a radical break from systems where the UV width is in one to one correspondence with different quark masses). We note this but introduce a UV cut off in the AdS space that at least seems to give a sensible description for quarks of any mass upto the cut off and removes the ultra-strong coupled regime. 

When a UV cut off is introduced one gains the additional ability to change the UV boundary condition of fields and, via Witten's multi-trace prescription, this allows us to include Nambu-Jona-Lasinio operators \cite{Nambu:1961tp} that can also trigger chiral symmetry breaking. We thus now have a system with many variables to explore. 

Thermal transition might be expected to occur when the black hole horizon grows with temperature to swallow more and more of the holographic radial direction including the tip of any U-shaped domain wall configurations. The preferred configuration is then two disconnected domain walls that fall into the black hole.  We briefly demonstrate this first order transition for domain wall configurations of fixed UV width. However, this is naive. In the presence of temperature and a UV cut off, we must be careful to compare between the \textit{same} theories at different temperatures. In particular we need theories with the same quark mass at the UV cut off, and we find that this is not captured by ordering the configurations by their width. A benefit of the domain wall method is that there is a holographic field on the domain wall that is directly dual to the quark mass and condensate. By solving for this fields profile we can label configurations by the quark mass. When we do this we find that the configurations re-assemble themselves to give a second order thermal, meson melting transition. 

Alternatively we can interpret our solutions using Witten's multi-trace prescription as the massless theory with a Nambu-Jona-Lasinio operator triggering chiral symmetry breaking. Here we must label configurations by the value of the NJL coupling $g$ and after doing so find in this case a first order chiral restoration transition.

Finally we can include confinement by compactifying the $x_4$ direction of the D5 branes. The low temperature behaviour has confinement and chiral symmetry breaking whilst above a first order transition to the deconfined phase the description returns to that of the description already described.

\section{I Domain Wall Configuration}

Let us first consider the basic domain wall set up. The gauge degrees of freedom are described by the geometry generated by $N_c$ D5 \cite{Itzhaki:1998dd}
(with $U=r/\alpha'$, $K= \frac{(2\pi)^{3/2}}{g_{YM}\sqrt{N}}$)
   \beq 
   {ds^2 \over \alpha'} =  K U(- dt^2 + dx_{1-4}^2)~~ +   \frac{1}{K U}\Big(d\rho^2 + \rho^2d\Omega_2^2 + dx_9^2\Big)  \eeq
   where $U^2 = \rho^2 + x_9^2$
   \beq
   e^\phi ={U \over K}, ~~~~~~ g_{YM}^2=(2\pi)^{3}g_s \alpha'   \eeq
   Note that in the 5+1d dual the gauge field is of energy dimension one so $1/g^2_{YM}$ has energy dimension two.  Here we see that $U$ has dimension one and the dilaton is dimensionless. 

Quark fields, initially on a 4+1 dimension defect in the 5+1 dimension gauge theory, are included via probe D7 branes.
The Dirac-Born-Infeld (DBI) action of a probe D7 brane describes its position in $x_9$ as a function of $\rho$ and $x_4$ is (up to angular factors)
\beq S_{D7} \sim \int d\rho\; e^{-\phi} (KU)\rho^2\sqrt{1+(\partial_\rho x_9)^2+{1 \over (KU)^2} (\partial_4 x_9)^2 } \label{d7act}\eeq 

\subsection{A Domain Wall Locus}

We now assume a solution $x_9=m$ with $m$ very large throughout the space except at particular values of $x_4$ where the sign of $m$ switches. At these $x_4$ positions $\partial_\rho x_9  \rightarrow \infty$. We now take the leading terms in the action in $\partial_\rho x_9$ and set it proportional to a delta function on the locus of the domain wall $x_4(\rho)$
\beq \partial_\rho x_9  = \left. {(KU)^{1/2} \over  (\partial_4 \rho)}\right|_{\rm locus} \delta(x_4-x_4(\rho)) ] \label{delta}
\eeq

\noindent where we have explicitly shown the Jacobian factor needed in the delta function.
We are left with an action for the position of the locus of the domain wall $x_4(\rho)$  
\beq S_{locus} \sim \int d\rho\; e^{-\phi} (KU)^{1/2} \rho^2 \sqrt{1+ (KU)^2 (\partial_\rho x_4)^2} \eeq which yields an equation of motion for the locus $x_4$ 
\beq \partial_\rho x_4 = {\pm \rho_m^{5/2}\over \rho \sqrt{\rho^5-\rho_m^5}}\eeq where $\rho_m$ is the conserved quantity which physically gives the minimum $\rho$ value that our solutions will reach. Integrating gives 
\beq x_4(\rho) = \pm{2 \over 5}  \arctan{\sqrt{\left({\rho \over \rho_m}\right)^5-1}}\eeq 

Thus these solutions are U-shaped configurations where $m=0$ on a 3+1 dimensional space with a holographic coordinate. The dual on this locus will describe the chiral fermions located thereon.

The U-shaped loci cup off at $\rho=\rho_m$ and then all have precisely the same width $2 \pi/5$ at large $\rho$. This is a  distinct behaviour from the configurations we found in the D3/probe D7 case in \cite{CruzRojas:2021pql} - there any asymptotic separation could be achieved. In fact in \cite{Evans:2021zzm} we did not notice this peculiarity because when a direction of the D5 is compactified to include confinement this behaviour ceases and arbitrarily wide configurations exits. There, this behaviour only manifests for heavy quarks, that were not our interest. This phenomena is a further odd behaviour for the D5/probe D7 system in the far UV. Note this behaviour suggests that the UV width of the configuration can not be taken to directly measure a UV parameter such as the quark mass. Our analysis of the DBI fields in the next section, which explicitly describe the quark mass,  confirms  this interpretation. Further at finite temperature, we will find that in a system at a given temperature there can be two configurations of equal UV width, but with different quark masses. Below we will sort the solutions by the quark mass or NJL interaction coupling rather than width.

A natural solution to the far UV behaviours is to simply include a UV cut-off at some fixed $\rho_{UV}$. The solutions then can take any width value in the UV up to a maximum of $2 \pi/5$ (corresponding to $\rho_m=0$). The zero width value is when $\rho_m=\rho_{UV}$. Presumably wider solutions are simply unstable to mutual attraction in the $x_4$ direction. Our next test of these solutions is whether they have sensible and consistent theories living on the loci. 

\subsection{B Domain Wall Theory}

The Lagrangian governing the fluctuations on the domain wall locus is given by inputting the delta function  (\ref{delta}) by hand into (\ref{d7act}). This enforces the geometry of the locus and the resulting domain wall theory is considered a fluctuation on the background U-shaped configuration (see \cite{CruzRojas:2021pql} for more details on the procedure). The action is
\beq S_{DW} \sim \int d\rho\; \rho^{5/2}\left(\partial_\rho x_4\right)\sqrt{1+{\cal A}(\partial_\rho x_9)^2+{1 \over (KU)^2} (\partial_\mu x_9)^2 },\eeq \\ with the function \beq {\cal A} = 1+ {1 \over (KU)^2 (\partial_\rho x_4)^2}\eeq encoding the $x_4$ dependence of the holographic fields. The equations of motion for the vacuum of the theory, 
\begin{align}
   {1 \over 2} \rho^{5/2}& \left(\partial_\rho x_4 \right) {\left({\partial{\cal A}\over \partial x_9}\right)(\partial_\rho x_9)^2 \over \sqrt{1 + {\cal A}(\partial_\rho x_9)^2}}\nonumber\\& - \partial_\rho\left({\rho^{5/2}\,\partial_\rho x_4 \,{\cal A}\,(\partial_\rho x_9) \over \sqrt{1 + {\cal A}(\partial_\rho x_9)^2}}\right) = 0,
\end{align}
 which shows that the $x_9(\rho)$ field has solutions with $x_9 = const$. It is natural to set this constant to the minimum radius $\rho_m$ that the U-shaped locus reaches to - the IR mass gap. The constant solution matches that seen in
 
  \begin{center}
        \includegraphics[width=0.9\linewidth]{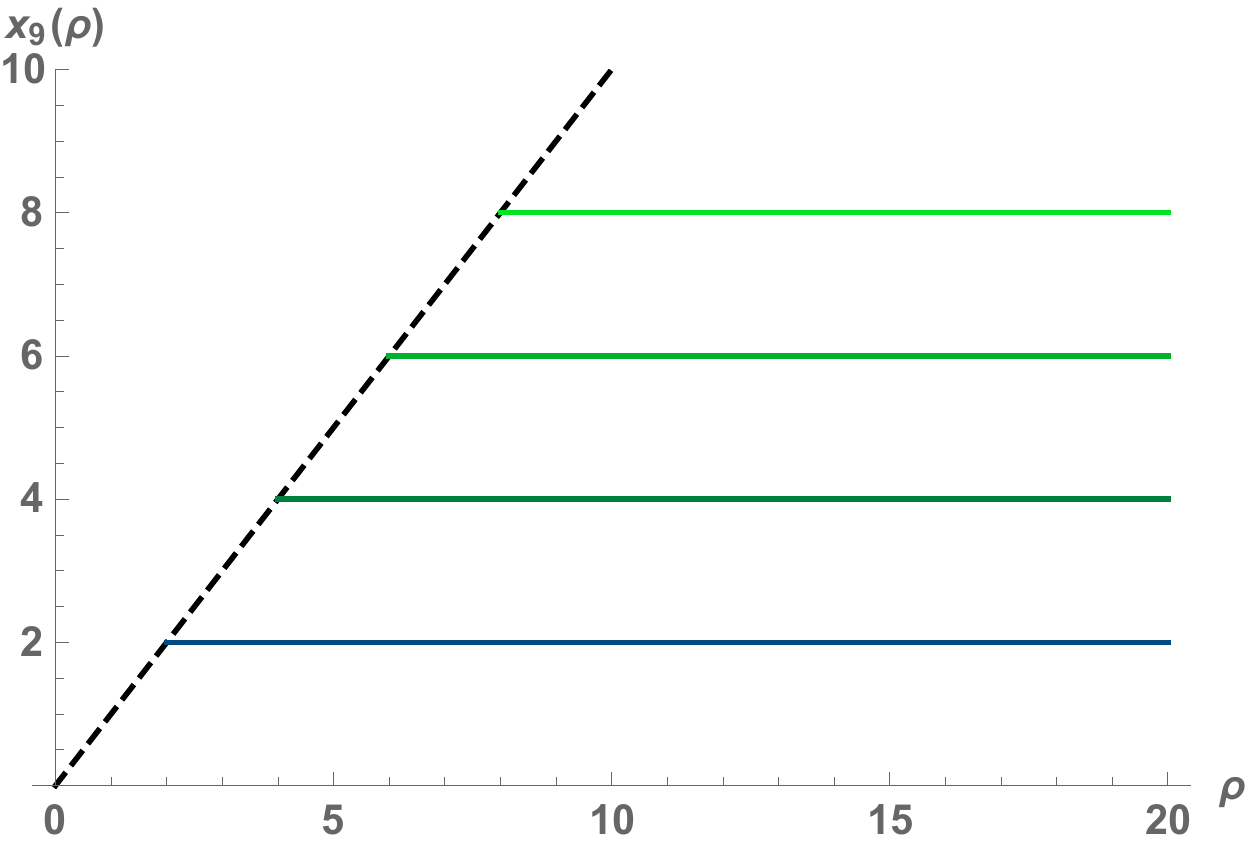}\\ \textit{Figure 1: Solutions for the profile of the $x_9$ field at zero temperature. The dashed line represents the IR boundary condition $x_9(\rho_{m})=\rho_{m}$. These configurations have different UV quark masses although they all tend to the same width in the far UV.  }
        \end{center}

 D3/probe D7 domain wall  configurations where there the fields on the domain walls are simple states with vanishing condensate and no renormalisation of the quark mass on their loci\cite{CruzRojas:2021pql}.  We plot these solutions in Figure 1 to display the IR boundary conditions and show that different $\rho_m$ domain walls lead to different UV masses.
 
 We stress again here that we have many configurations with the same far UV width of $2\pi/5$ but with different IR masses $\rho_m$. The holographic field on the locus dual to the quark mass provides supporting solutions with different values of the UV mass. The UV width can not be taken to measure the quark mass but the holographic field on the locus does allow a clean interpretation of the configurations.

  In the next section, we will repeat the above construction in the background of a black fivebrane to desribe the finite temperature behaviour of the system.

\section{II Finite T - meson melting transition}
The supergravity solution for the near horizon geometry around a stack of black fivebranes with $N_c$ units of RR 6-form flux is ($U=r/\alpha'$, $K= \frac{(2\pi)^{3/2}}{g_{YM}\sqrt{N}}$) \cite{Itzhaki:1998dd} 
   \beq  \label{metricT}
   {ds^2 \over \alpha'} =  K U(- h dt^2 + dx_{1-4}^2)~~ +   \frac{1}{K U}\Big({1 \over h}dU^2 + U^2d\Omega_3^2\Big)  \eeq
   where 
   \beq
   h(U) = 1- {U_0^2 \over U^2}\eeq \beq
   e^\phi ={U \over K}, ~~~~~~ g_{YM}^2=(2\pi)^{3}g_s \alpha' \eeq
  As before, the dual gauge field is of energy dimension one and  $1/g^2_{YM}$ has dimension two.  Here we see that the raidal direction $U$ has dimension one and the dilaton is dimensionless. $U_0$ is the position of the horizon and its value is proportional to the temperature.
  
  It is helpful to perform a coordinate transformation to make the presence of a flat 4-plane transverse to the horizon manifest, as naturally seen by the embedded probe D7 branes. This involves a transformation from the $U$ coordinates to a dimensionless set of $v$ coordinates 
   \begin{equation} {U\over U_0} = {1+v^2\over 2v}. \end{equation} 
   $U_0$ has dimension one and, as the position of the horizon in the geometry, encodes the temperature of the thermal state in the dual gauge theory. Note that $v$ is dimensionless and that the black hole horizon always lies at $v=1$. To introduce another scale to compare to $U_0(T)$, and to remove the strongly coupled far UV, we will introduce a UV cut off, $v_{UV}$ and work at $1 \leq v \leq v_{UV}$. In the spirit of the discussion in \cite{Evans:2021zzm}, one can imagine the UV cut off to correspond to the scale where an asymptotically free gauge theory such as QCD moves from UV weak coupling to IR strong coupling. One would only expect a weakly coupled gravity dual below this scale. Since $v=1$ independently of $T$ in the $v$ coordinates we will vary $v_{UV}$ to adjust the ratio of these two scales. Changing the ratio
   can be interpreted either as moving the cut off at fixed $T$ or, as we will choose, varying $T$ at fixed cut off.

   In the $v$ coordinates, the metric becomes
   \beq ds^2 = G_x (-h(v)dt^2 + (dx^{1-4})^2)+G_v(d\rho^2 + \rho^2 d\Omega_2^2 + dL^2), \eeq
with metric factors \beq G_x= K U_0 {v^2+1 \over 2 v}, \hspace{1cm}  h(v) = 1 - \left( {2 v \over v^2 +1} \right)^2,\eeq  
\beq G_v =  {U_0\over K}{1+ v^2\over 2v^3}, \hspace{1.7cm} e^{-\phi} = {K \over U_0} {2v \over 1+v^2}. \eeq 
The position of probe D7s in this geometry is described by the DBI action for the embedding $L(\rho,x_4)$
 \begin{align} S_{D7} \sim \int d^8x\; h(v)& e^{-\phi} G_x^{5/2}G_v^{3/2}\rho^2\nonumber \\\times&\sqrt{1+(\partial_\rho L)^2 + {G_v \over G_x}(\partial_4 L)^2}.\end{align}
 Here again we assume a solution $L = m$ which sharply changes sign along a contour $x_4(\rho)$. Taking $\partial_4 L$ of the form of a large number times a delta function on the domain wall locus $x_4(v)$ dimensionally reduces the action of the D7 embedding to \begin{align} S_{locus}\sim\int d^7x\; h(\rho)& e^{-\phi(\rho)}G_{x,l}^2 G_{v,l}^{3/2}\rho^2\nonumber\\ \times& \sqrt{1 + {G_{x,l}\over G_{v,l}}(\partial_\rho x_4)^2},\end{align}
 where the quantities $G_{x,l},~ G_{v,l}$ are the metric factors on the locus (in this case, where $v = \rho$). This action gives the equation of motion for the domain wall locus 
 \beq  \partial_\rho x_4 = {\pm {\cal C} \;G_{v,l}^{1/2} \over G_{x,l}^{1/2}\sqrt{h^2 e^{-2\phi}G_{x,l}^5 G_{v,l}^2\rho^4-{\cal C}^2}}.\eeq
 which can be integrated numerically to solve for the position of the domain walls. These loci share similarities with the domain wall loci in the non-thermal geometry. Like (8),
 the numerical solutions from (21) are U-shaped and cup off at some $\rho_{min}$. The connected solutions have an upper maximum width. There also exist disconnected solutions, where the tip of the domain wall pair has been swallowed by the black brane horizon, and the locus splits into two flat pieces that are effectively screened from one another. This disconnected solution naturally exists at all UV domain wall separations and which the system will prefer is determined by their respective free energies. The fluctuations about the disconnected solutions are quasi-normal modes \cite{Hoyos-Badajoz:2006dzi}, indicating that mesons in the theory will dissipate into the thermal plasma, and thus we are in a melted phase. The following section will examine theories that live on the connected loci and the phase transitions they exhibit.
 
 \subsection{A Thermal transition with respect to width.}
 
 To begin to orientate ourselves to the solutions at different $U_0(T)/\Lambda_{UV}$ values let us be naive and treat the UV width of a configuration at $\Lambda_{UV}$ as a parameter. We stress these will not turn out to be configurations that share any UV Lagrangian parameter though. 
 
 Thus for each choice of $v_{UV}$ we seek solutions for the U-shaped domain wall loci that share a particular $x_4$ width. Typically at large $v_{UV}$ (low temperature) there are three solutions, as depicted in Figure 2: two connected solutions and one disconnected set of domain walls.  For small choices of $v_{UV}$ (high temperature) only disconnected solutions exist for the same width - the horizon grows out towards the UV boundary and ``eats" the connected solutions.

 \begin{center}
\includegraphics[width=0.9\linewidth]{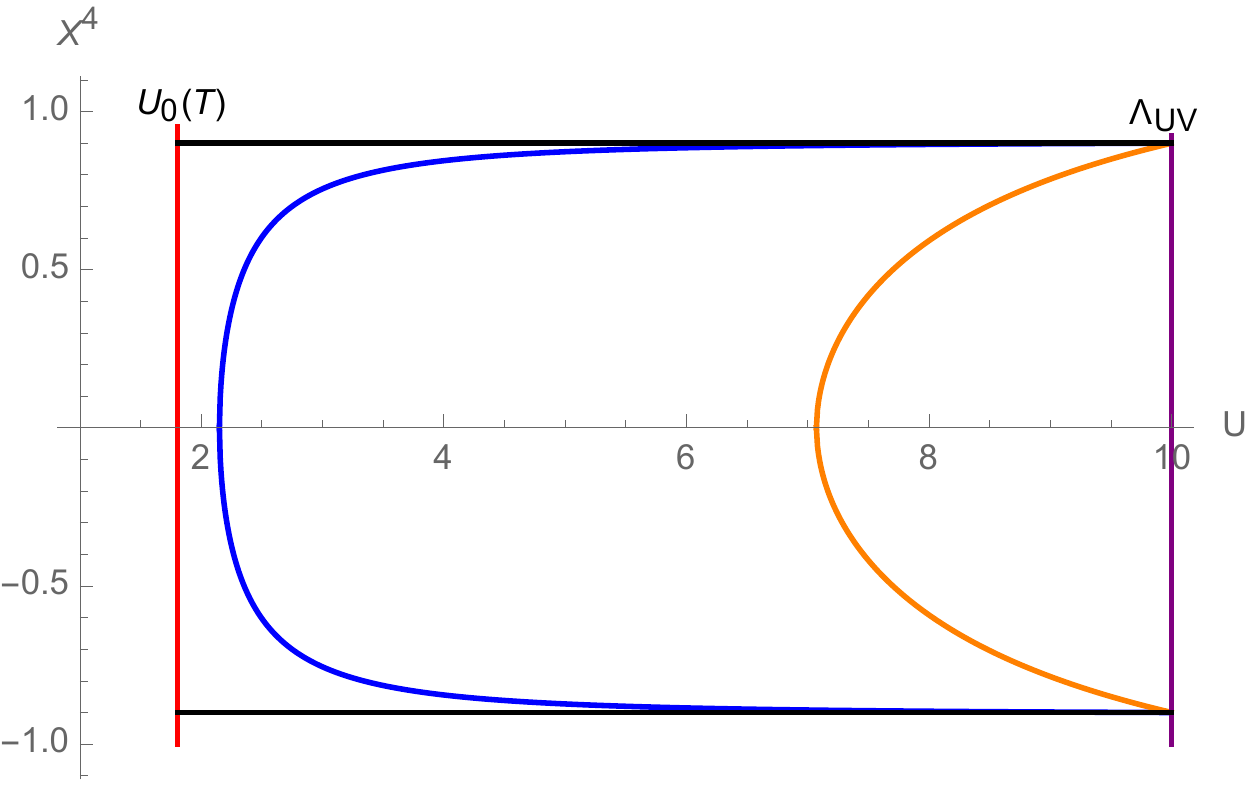}
\textit{Figure 2: A cartoon of three domain wall loci solutions at a fixed temperature that share the same UV width.}
\end{center}

\begin{center}
\includegraphics[width=0.9\linewidth]{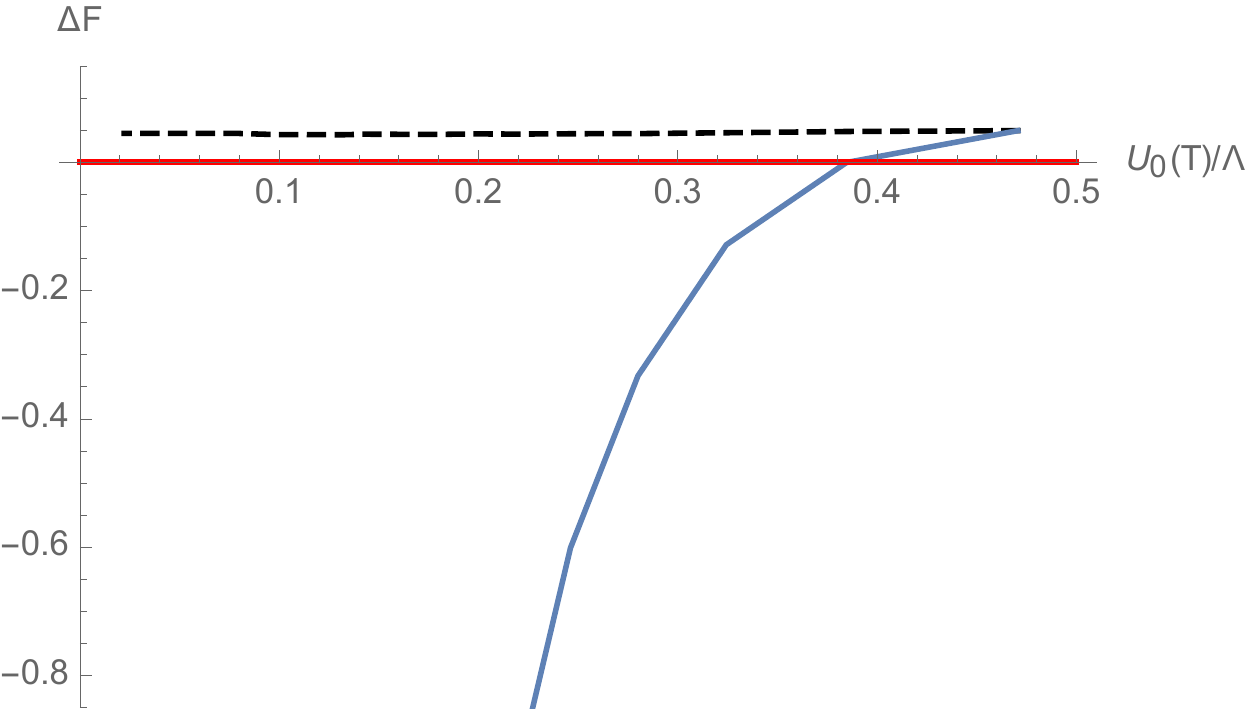}
\textit{Figure 3: The free energy of the three different domain wall configurations of fixed width - two Us (blue and black dotted) and the disconnected (red) case - against $U_0$ or $T$. We see a first order transition with respect to temperature, here with $w=0.9$}
\end{center}

  The free energy,  in the absence of any other thermodynamic variables (such as chemical potential), is simply \beq F \simeq -S_{locus},\eeq
Here we consider  a free energy that is a function of the ratio of the UV width and the temperature $F = F(W,U_0(T)/\Lambda_{UV})$. For the disconnected solutions, the free energy is independent of the width, since they are decoupled. At a given temperature $F_{disconnected}$ is just a constant, that is solely dependent on the UV boundary, and can be subtracted from every solution to remove the cutoff dependence. 

Calculating these actions for this system we find the results shown in Figure 3 suggesting a first order phase transition. The connected solution is favoured at small temperature; the disconnected solution at large temperature; the second connected solution that lies closest to the horizon has the highest energy of all the configurations and must correspond to the effective potential maxima between the two minima that interchange at the transition.

 As we will now show though, this analysis is not just naive but gets the order of the meson melting transition wrong. To fully understand the configurations we should require them to have the same UV Lagrangian parameters. In systems such as the D3/probe D5 ${\bar D5}$ system this identification is hard because the configuration does not provide a holographic field dual to the quark mass and condensate. In the domain wall system though the field dual to the 4+1 dimension quark condensate can still fluctuate on the domain wall and provides a field to enable this identification as we have argued in \cite{CruzRojas:2021pql}. We will pursue this line of reasoning in the next section.

 \subsection{B Domain Wall Theory at Finite T}
 Now we turn to the holographic theory that lives on the domain wall loci, and set out the prescription by which we can identify the quark mass for a given domain wall locus. In the thermal geometry one has, after restricting the DBI fields to the domain wall locus,
 \begin{align}
     S_{DW}\sim \int d^6x \int_{1}^{v_{\Lambda}}\!\!\!d\rho  \;h(v)e^{-\phi(v)}{G_x^{5/2}(v)G_v^{3/2}(v) \over G_{v}^{1/2}(\rho)}\nonumber\\\times\left(\partial_\rho X^4\right)  \rho^2 \sqrt{1 + \Gamma(\partial_\rho L)^2}.
 \end{align} with $\Gamma = 1+ {G_v(v)\over G_x(v)(\partial_\rho X^4)^2 }$. The equations of motion for $L$ then follow straight forwardly.  The field $L$'s UV asymptotics gives information about the mass and chiral condensate in the theory with large $v$ solution
 \beq L\sim \Tilde{m}+{\Tilde{c}\over v^3}.\eeq 
 However, it is noted that $L$ is in dimensionless units, and as such does not represent the physical, dimensionful quantities. For the case of the mass, we have \beq {m \over \Lambda} = {1+\Tilde{m}^2 \over \Tilde{m}}{v_\Lambda\over 1+v_\Lambda ^2}.\eeq The condensate is proportional to $\tilde{c}$ \cite{Evans:2021zzm} and, assuming the cut off $v_{UV}\gg 1$, we have
 \beq {\langle\bar{q} q \rangle \over \Lambda^3} \propto {\Tilde{c} \over v_\Lambda^3}.\eeq 
 
 In the IR, on the U-shaped loci, we again assume $L(\rho_{min})=\rho_{min}$ and $L'(\rho_{min})=0$ so that the IR mass gap is consistent. One then shoots out to determine the UV mass and condensate values. 
 
 With these identifications of the UV parameters, it is possible (numerically) to study the system at varying temperature and categorise the observed phenomena. In practice in any given geometry with horizon at $v=1$ and some choice of UV cut off the easiest variable to 
 
 \begin{center}
        \includegraphics[width=0.9\linewidth]{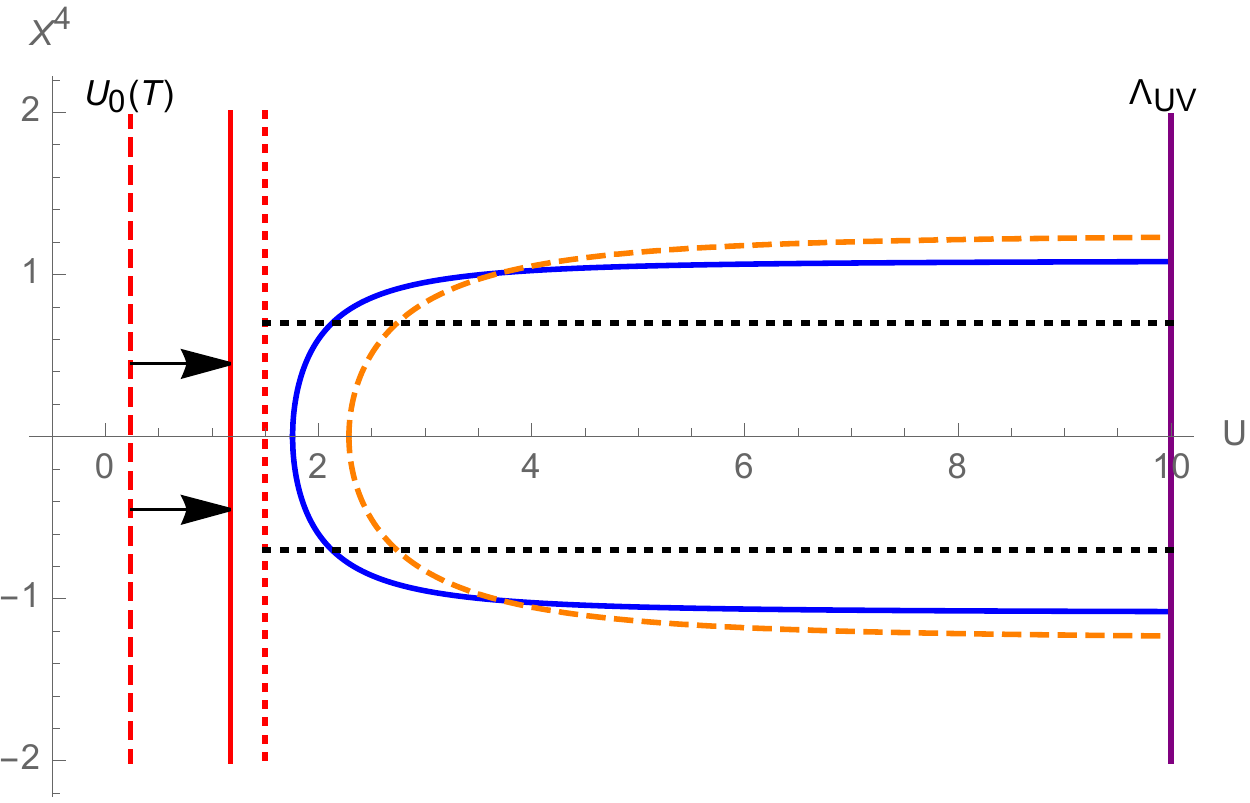}\\ \textit{Figure 4: A cartoon showing the evolution of a Domain Wall system with constant $m/\Lambda$ under an increase of temperature T (red, dashed) to $T'>T$ (red, solid), the connected locus begins to ``square off''. Upon increase from $T'$ to $T_c>T'$ (red, dotted) the tip of the locus falls into the horizon and the locus becomes disconnected (dotted, black).}
        \end{center} 
 \begin{center}
        \includegraphics[width=0.9\linewidth]{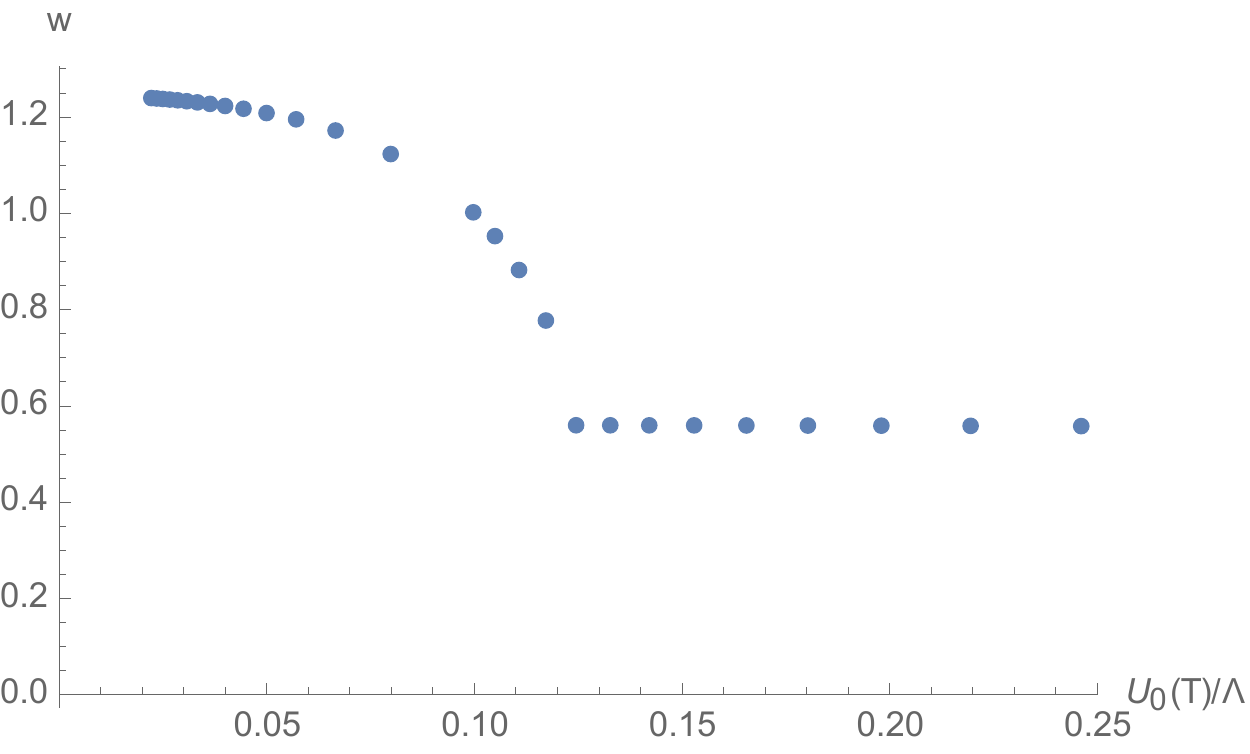}\\ \textit{Figure 5: UV widths of the domain wall configurations of mass $m/\Lambda =0.15$. At a critical temperature $U_0(T_c)/\Lambda\sim 0.08$ The U shaped loci drop into the horizon and split into two separate screened pieces, marking the second order transition.}
    \end{center}
 
 change remains the width via the choice of $\rho_{min}$. On each such configuration we then solve the domain wall theory to identify the UV mass. From a sequence of such tabulated results one can identify a particular mass for a particular $U_0(T)/\Lambda_{UV}$.

 Thus we now discuss theories at fixed $m/\Lambda_{UV}$. When the temperature lies far below $\rho_{min}$ the IR mass scale of the U-shaped loci then the width of the locus at the UV cut off maps to the quark mass. As the temperature rises though the domain wall loci that describe an equal quark mass begin to shift and the UV width shrinks. We sketch this behaviour in Figure 4 and show data for the width against temperature in Figure 5.

 As one raises the temperature further the U-shaped loci begin to ``square off'' and the minimum point approaches the horizon. In the width versus temperature plot, Figure 4, we can see that the width of these configurations 
 
 \begin{center}
         \includegraphics[width=0.9
    \linewidth]{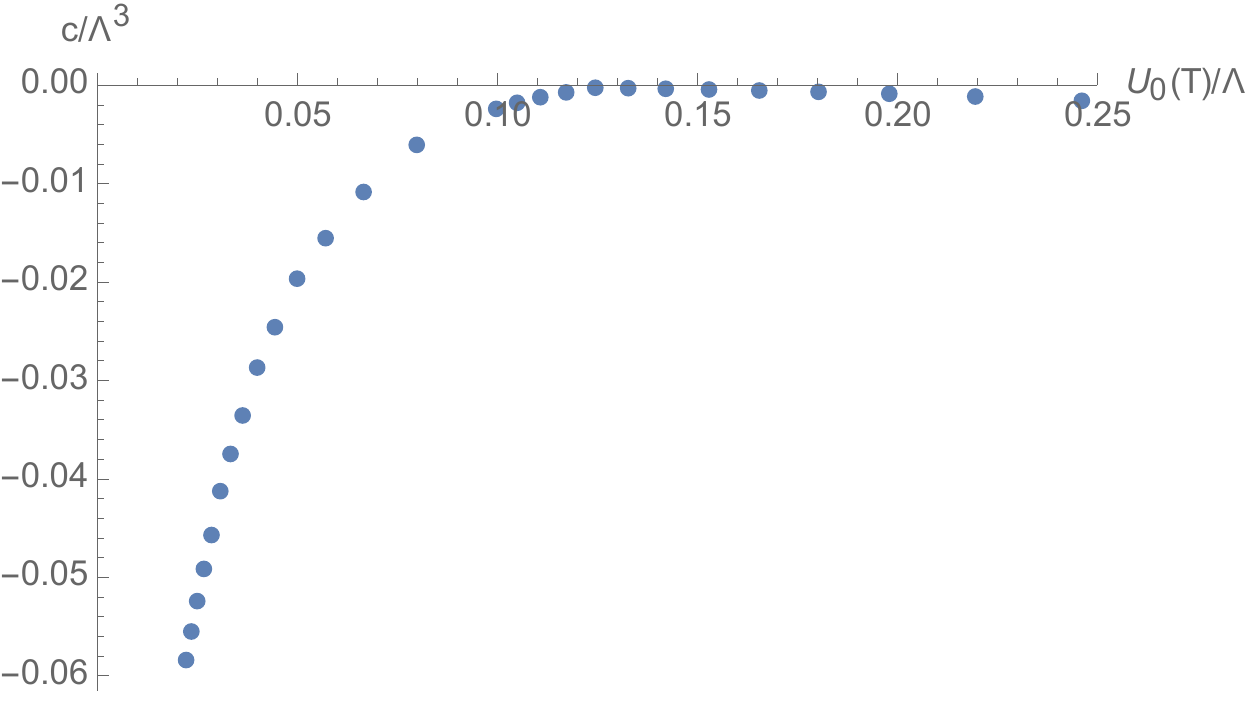} \\\textit{Figure 6: The corresponding condensate for configurations with $m/\Lambda = 0.15$, at the critical temperature the condensate vanishes, indicating the solutions have $\partial_\rho L =0$.}
    \end{center}

\begin{center}
        \includegraphics[width=0.9\linewidth]{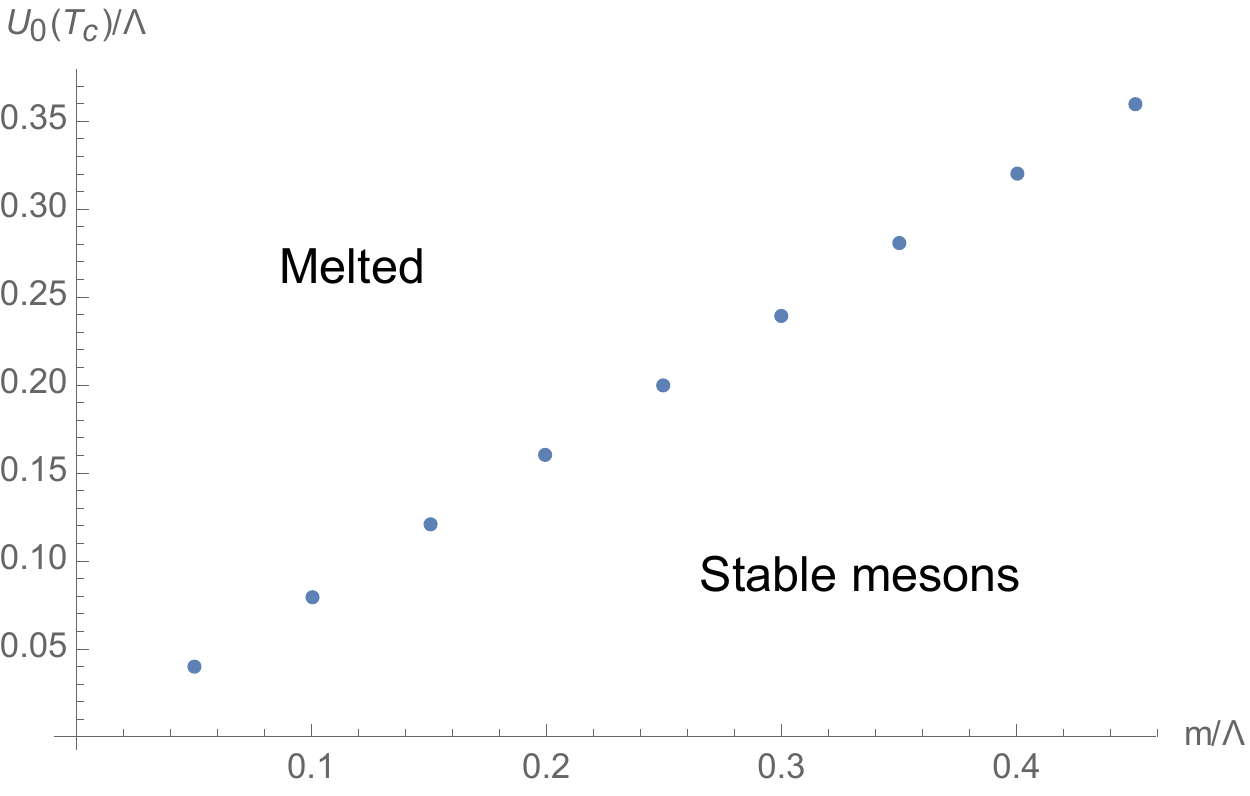}\\ \textit{Figure 7: Variation in critical temperature $U_0(T_c)/\Lambda$ with changes in the quark mass $m/\Lambda$}\\  
    \end{center}
 
 freezes at a fixed value. Our interpretation is that these solutions, which essentially live parallel to the black hole horizon until kinking to a straight line in $v$, are really the systems best attempt to describe two disconnected branes. This is a second order transition to the meson melted phase of the system. 

We stress that when ordering the domain wall loci by equal quark mass, the first order thermal transition seen with width changes to a continuous second order transition. This is achieved by the width of the U-shaped loci changing smoothly, to resort the configurations naively ordered by width, at high temperature to allow this smooth behaviour. To emphasise the second order nature of the transition we plot the quark condensate against temperature in Figure 6 to show that it changes continuously at the transition.  As far as we know, this is the first example of a second order meson melting transition in holography.
 
At T=0 the system reproduces the flat solutions for the domain wall loci field $x_9$ of the non-thermal geometry in the section IB, where $m_{IR}=m_{UV}$. Intuition suggests that the system will exhibit a transition when $T>m$, which we indeed find to be the case. We plot how the critical temperature varies with the quark mass in Figure 7 - it shows that there is direct proportionality between $U_0(T_c)$ and $m$. Thus for massless quarks, the phase transition occurs for any $T>0$.

\subsection{C NJL interpretation}

Witten's multi-trace operator prescription \cite{Witten:2001ua} teaches us that where a solution such as those we have discussed has a UV quark mass there are two interpretations. Either one has a quark mass arising form a bare lagrangian term, 
or something like a Nambu-Jona-lasinio (NJL)  interaction \cite{Nambu:1961tp} that dynamically generates a quark mass. We examine this case further here. An effective four fermion interaction 
\beq\Delta {\cal L} = {g^2 \over \Lambda_{UV}^2} \Bar{\psi}_L\psi_R\, \Bar{\psi}_R \psi_L \xrightarrow{\langle \Bar{\psi}_L\psi_R \rangle}m\Bar{\psi}_R \psi_L,\eeq
 upon condensation of the $\Bar{\psi}_L\psi_R$ operator, generates an effective mass term for the quarks. This four fermion interaction is a double trace operator and we can follow the prescription in \cite{Witten:2001ua}. They arise as boundary conditions on the fields in the supergravity theory, and for the case of holographic NJL interactions, this was explored in \cite{Evans:2016yas}. It was found that one could add terms to the Lagrangian 
to impose \beq {g^2 \over \Lambda_{UV}^2} \langle \Bar{\psi}_L\psi_R \rangle =m\eeq at the classical level. 
 
 Following suit, we can reinterpret each configuration in our analysis above with $m$ the quark mass and $\langle \Bar{\psi}_L\psi_R \rangle$ the quark condensate. The mass can be interpreted as dynamically generated and we can calculate the NJL coupling $g^2$ from the asymptotic values of the holographic field $L$, with \beq g^2 = {1+\Tilde{m}^2 \over \Tilde{m}\Tilde{c}}{v_\Lambda^4 \over 1+v_\Lambda ^2}. \eeq 
 
 Now we must again sort our U-shaped configurations at each $U_0(T)/\Lambda_{UV}$ but by $g^2$ rather than $m$ - it is $g^2$ that is now defining the theory at the cut off. In this case we find that to each value of $g^2$ there are three configurations: two U-shaped configurations and one flat solution that has fallen into the horizon. This behaviour is shown directly in Figure 8. Here the behaviour we saw at fixed width is recovered in contrast to that at fixed mass where there was just a single U-shaped configuration for a given mass value. 
 
 \begin{center}
        \includegraphics[width=0.9\linewidth]{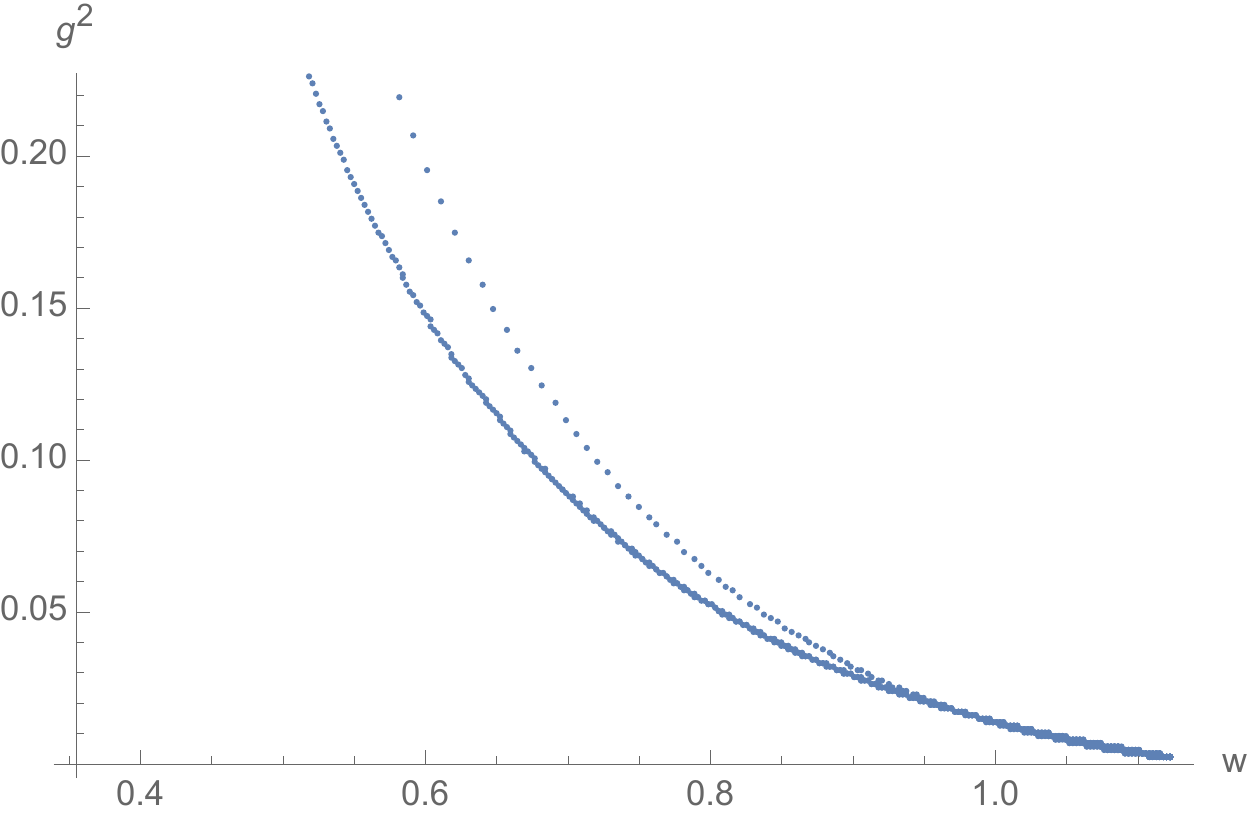}\\ \textit{Figure 8: The two U-shaped solutions of different widths (each with the same NJL coupling $g^2$) shown for a temperature of $U_0(T)/\Lambda \sim 0.22$ }
        \end{center}
        
              \begin{center}
        \includegraphics[width=0.9\linewidth]{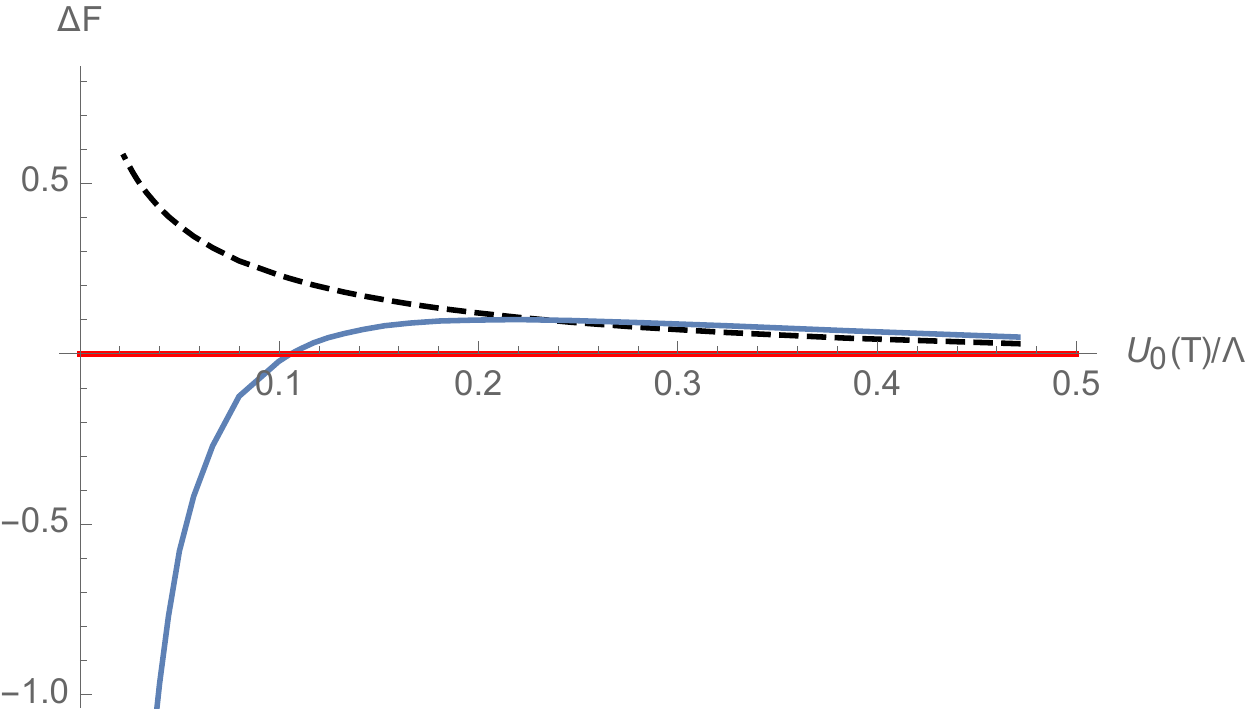}\\ \textit{Figure 9: The free energy of the three solutions (two U-shaped and flat) for $g^2=5$ against temperature showing the first order transition. }
        \end{center}

 As expected where there are three solutions, i.e. three stationary points of the action (and hence free energy) a first order phase transition is observed. One can imagine a system with a temperature dependent potential has three extrema, and is bounded from below. It must have two local minima, one local maxima, and one (or at the transition temperature two) global minima. The U-shaped configurations in this system represent one of the local minima, and the local maxima, with the flat locus representing the other local minimum. As the temperature changes and the two local minima become equal the system will undergo a discontinuous, first order phase transition. This is explicitly realised in Figure 9 where 
  the transition is shown for $g^2=5$ and the phase transition occurs at a critical temperature of $U_0(T_c)/\Lambda \sim 0.1$. 

  In Figure 10 we map the critical temperature of the phase transition as a function of $g^2$. Note here the transition is a chiral restoration transition as well as a meson melting transition.

    \begin{center}
        \includegraphics[width=0.9\linewidth]{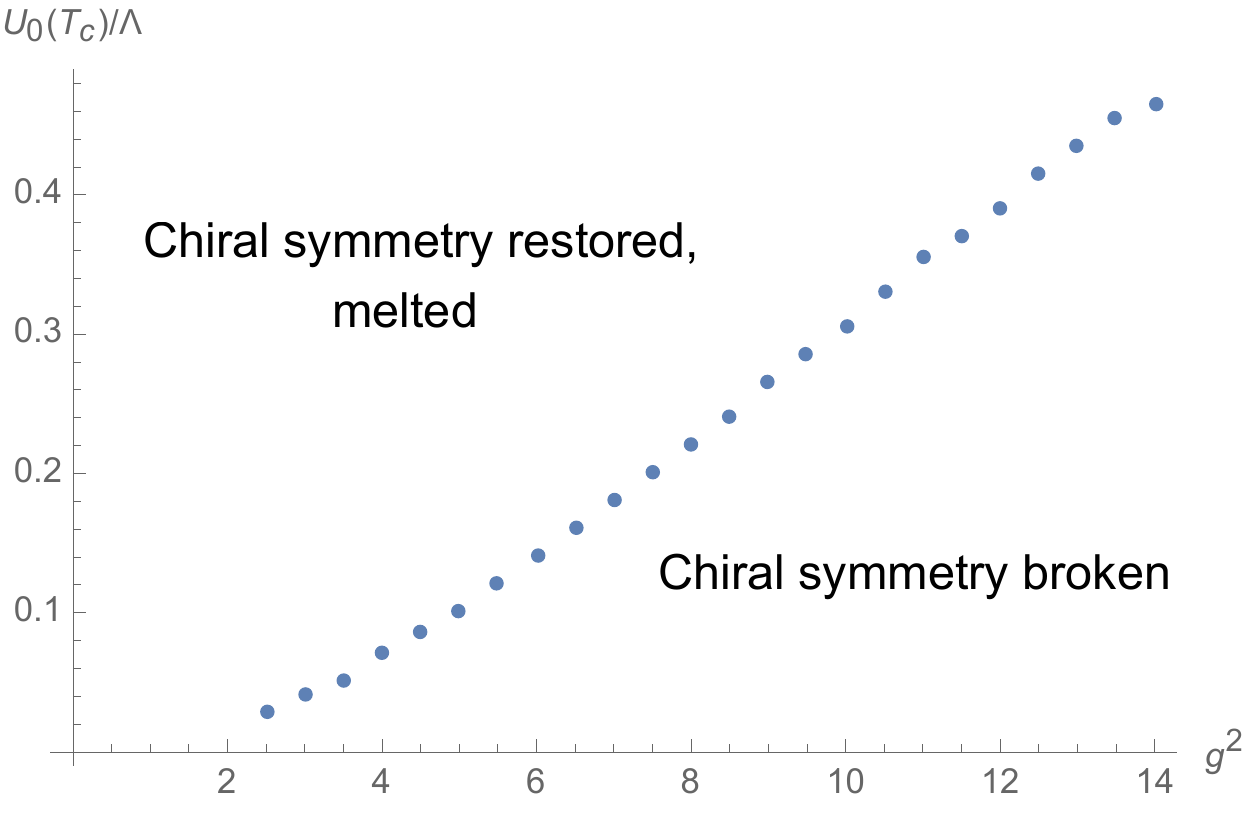}\\ \textit{Figure 10: Critical Temperature, $U_0(T_c)/\Lambda$ of the first order phase transition with respect to $g^2$. }
        \end{center}

\subsection{D Confining Geometry}
Finally we can consider the D5/probe D7 system with a compact $x_4$ direction. Compactifying the D5 introduces a confinement scale in the gauge dynamics. This system was studied in  \cite{Evans:2021zzm} where the confinement was shown to lead to chiral symmetry breaking in the domain wall theory. The confining geometry is given by \cite{Horowitz:1998ha}
 \beq  \label{metricT}
   {ds^2 \over \alpha'} =  K U(-dt^2 + dx_{1-3}^2 + h dx_4^2)~~ +   \frac{1}{K U}\Big({1 \over h}dU^2 + U^2d\Omega_3^2\Big)  \eeq
\beq
   h(U) = 1- {U_C^2 \over U^2}\eeq 
Directly comparing to (\ref{metricT}) one observes the usual story that the free energy switches from preferring the confined geometry at low $T$ to the finite temperature deconfined geometry when $U_0=U_C$. This means that our analysis above corresponds to that of the high temperature phase of this system. The low temperature phase is that described in \cite{Evans:2021zzm}. In practice this simply means that in our phase diagrams in Figures 7 and 10 one should draw a horizontal line across at $U_0=U_C$ and below this line the model is confined. Note that for large $g$ there remains chiral symmetry breaking even above the deconfinement transition separating the phenomena.

\section{III Summary}

The domain wall method to isolate chiral fermions should be a useful technique to build holographic models of QCD and wider chiral gauge theories.

In this  paper we have studied the thermal phase transitions in the D5/probe D7 system. Domain wall configurations are generically U-shaped with two domain walls meeting. We have studied the holographic field on the domain wall dual to the quark mass and condensate. In the presence of temperature this allows us to classify domain wall solutions by mass or NJL coupling. 

Our analysis shows that in the basic D5/probe D7 system the U-shaped configurations describe massive quarks. The meson melting transition, as a black hole horizon grows into the bulk, occurs when the tip of the U enters the horizon leaving two disconnected walls. We have shown this transition, at fixed mass, is second order.

In the NJL interpretation where one fixes the NJL coupling at the UV cut off, the chiral restoration transition is first order.

These configurations can also describe the high temperature deconfined theory above a first order transition from a geometry with a compact $x_4$ direction. The low energy theory below that transition has confinement and chiral symmetry breaking. 

In the future we hope to develop the technology described here to also include finite density. We should then be able to generate descriptions of the QCD phase diagram, of heavy ion collisions and of neutron star interiors. \\\newline
 \noindent {\bf Acknowledgements:} The authors would like to thank Andreas Schmitt and Aaron Poole for fruitful discussions. NEs work was supported by the STFC consolidated grants ST/P000711/1 and ST/T000775/1. JMs work was supported by an STFC studentship.

\end{document}